\documentclass{sf2a-conf2017}
\usepackage{graphicx}
\usepackage{hyperref}
\usepackage[]{natbib}  
\usepackage{epstopdf}
\usepackage[font=small,labelfont=bf]{caption}

\def\BibTeX{{\rm B\kern-.05em{\sc i\kern-.025em b}\kern-.08em
    T\kern-.1667em\lower.7ex\hbox{E}\kern-.125emX}}
\bibpunct{(}{)}{;}{a}{}{,}  

\begin{document}

\TitreGlobal{SF2A 2017}


\title{Hyperfine Structure and Abundances of Heavy Elements in 68 Tauri (HD 27962)}

\runningtitle{Hyperfine Structure and Abundances of Heavy Elements in 68 Tauri}

\author{S. Martinet}\address{Sebastien.Martinet@univ-grenoble-alpes.fr}

\author{R. Monier}\address{Richard.Monier@obspm.fr}




\setcounter{page}{237}


\maketitle


\begin{abstract}
  HD 27962, also known as 68 Tauri, is a Chemically Peculiar Am star member of the Hyades Open Cluster in the local arm of the Galaxy.
We have modeled the high resolution SOPHIE (R=75000) spectrum of 68 Tauri using updated model atmosphere and spectrum synthesis to derive chemical abundances in its atmosphere.
In particular, we have studied the effect of the inclusion of Hyperfine Structure of various Baryum isotopes on the determination of the Baryum abundance in 68 Tauri.
We have also derived new abundances using updated accurate atomic parameters retrieved from the NIST database. 
\end{abstract}



\section{Introduction}

68 Tauri (HD 27962) is the hottest and most massive member of the Hyades open cluster (age about 700 Myrs). Previous abundance analyses of HD 27962 have revealed a distinct underabundance of scandium and overabundances of the iron-peak and heavy elements which prompted to reclassify this early A star as an Am star. The last abundance analysis dates back from 2003 \citep{2003A&A...406..987P}. It seems therefore justified to redetermine and expand the chemical composition of this interesting object using updated atomic data.
In particular we have included the hyperfine structure for several lines. We present here the results for one line of Ba II and discuss the revision of the baryum abundance and other abundances in 68 Tauri.

\section{Abundance Determinations}
  
\subsection{Model Atmosphere and Spectrum Synthesis}

We used the observed Str\"omgren photometry of 68 Tauri retrieved from SIMBAD and the UVBYBETA code of T.T.Moon (1985) to determine an effective temperature of $9025\pm200$ K and a surface gravity log(g)=3.95$\pm$0.25 dex for 68 Tauri.
We used these parameters to compute a 72 layers plane parallel model atmosphere with the ATLAS9 code (Kurucz, 1992) assuming Local Thermodynamical Equilibrium, Hydrostatic Equilibrium and Radiative Equilibrium.


We used Hubeny's code SYNSPEC49 (1992) to compute a grid of synthetic spectra to model the observed spectrum of 68 Tauri.
We first computed a synthetic spectrum adopting solar abundances as a first iteration and then altered the abundances in order to reproduce the line profiles of selected lines with accurate atomic parameters.

\subsection{Baryum Hyperfine Structure}

We have replaced the single line~$\lambda~5853.675$\AA~of Ba II extracted from the NIST Atomic Spectra Database with the Hfs of the 5 major isotopes of 
Baryum as calculated by McWilliam(1998). We used a solar isotopic mixture
to compute the grid of synthetic spectra.
We find a large difference on the Baryum abundance when including the full hyperfine structure. 
For the 5853.675\AA~line shown Figure 1, the inclusion of Hfs yields a Baryum abundance lower by 0.4 dex than when ignoring Hfs.


\begin{center}

\includegraphics[width=0.75\linewidth]{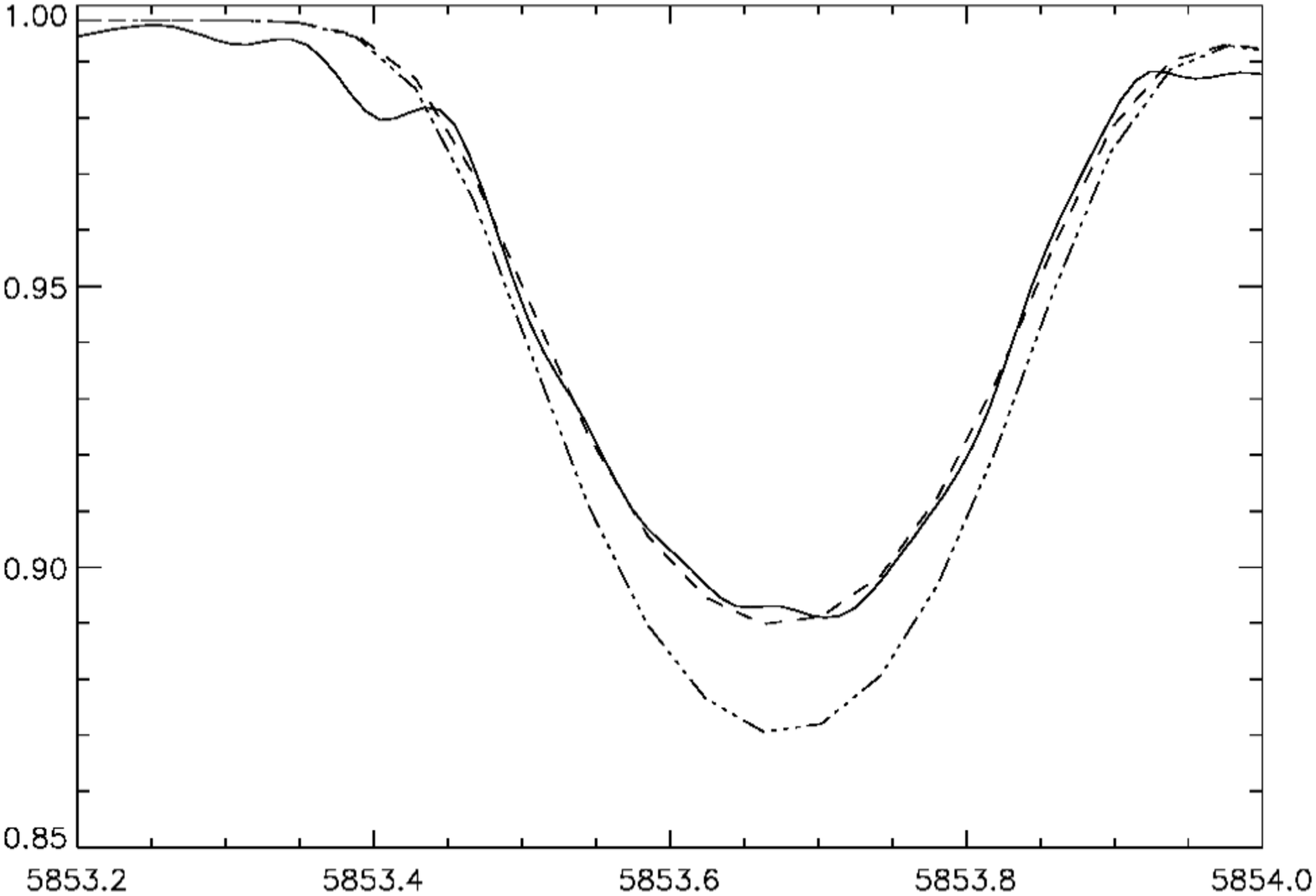}
\captionof{figure}{The effect of including Hfs on the line profile of the 5853.675~\AA~line of Ba II \textit{(observed: thick line, synthetic spectra: dashed line)}}
 \begin{picture}(300,0)
  \put(-5,185){\makebox(0,0)[r]{\small{Normalised Flux}}}
  \put(200,43){\makebox(0,0)[r]{\small{Wavelength~$\lambda$~(in \AA{})}}}
  \put(358,117){\makebox(0,0)[r]{\small{$\longleftarrow$ Without Hyperfine Structure}}}
  \put(352,150){\makebox(0,0)[r]{\small{$\longleftarrow$ With Hyperfine Structure}}}
 \end{picture}

\end{center}

\subsection{Abundances in the Atmosphere of 68 Tauri}

For iron, we have determined abundances for each lines of Fe II and then computed a weighted mean according to the quality grades assigned to each transition in NIST. 
\begin{center}

\includegraphics[width=0.75\linewidth]{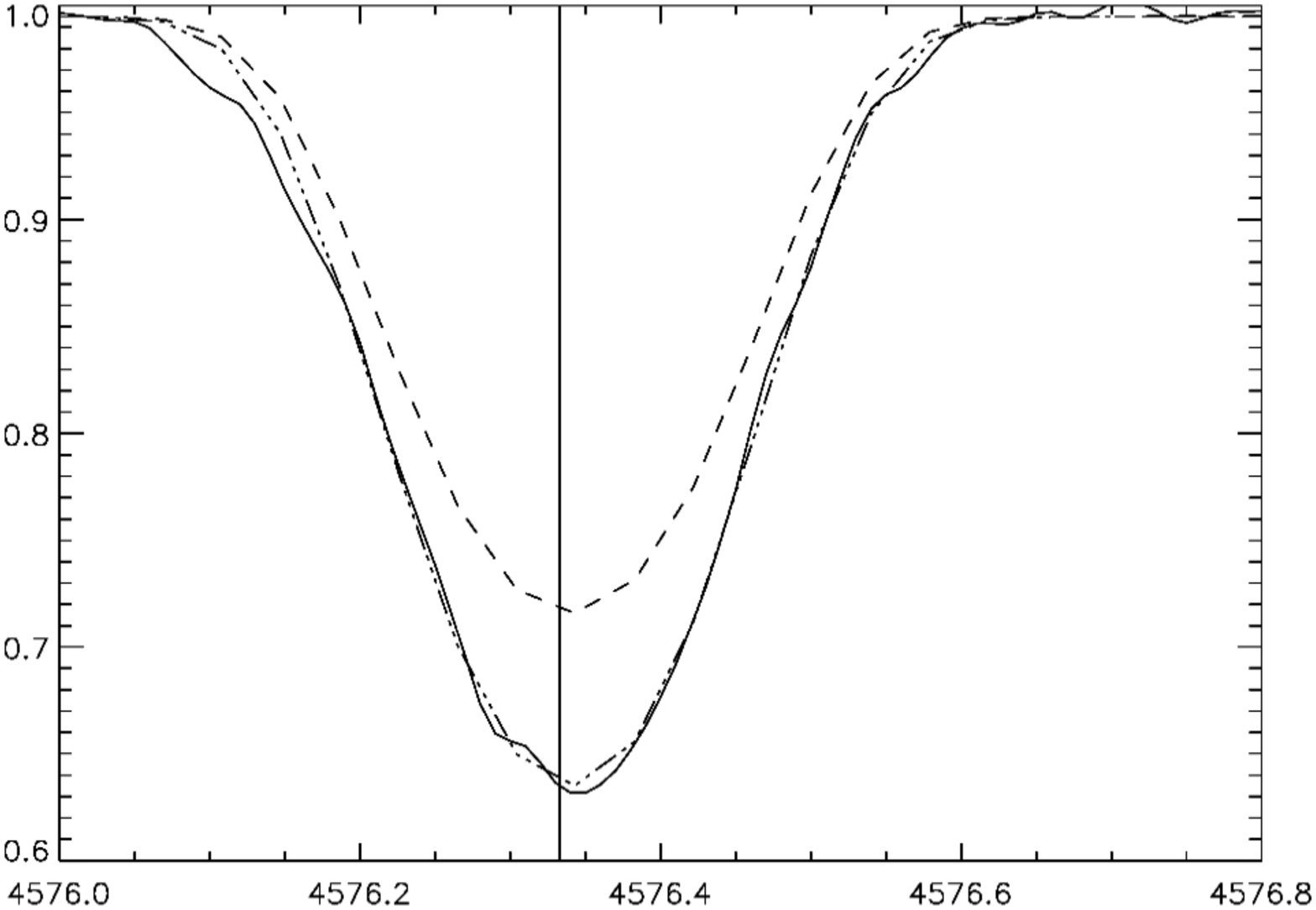}
\captionof{figure}{Determination of Iron abundance for the 4576.33\AA~Fe II line \textit{(observed: thick line, synthetic spectra: dashed line)}}
\begin{picture}(300,0)
  \put(-5,185){\makebox(0,0)[r]{\small{Normalised Flux}}}
  \put(192,45){\makebox(0,0)[r]{\small{Wavelength~$\lambda$~(in \AA)}}}
\put(186,294){\makebox(0,0)[r]{\small{$\lambda_{NIST}=4576.333$\AA}}}
\put(208,150){\makebox(0,0)[r]{\small{$\longleftarrow$ 1.00$\odot$}}}
\put(203,100){\makebox(0,0)[r]{\small{$\longleftarrow$ 2.03$\odot$}}}
\end{picture}

\end{center}


We then applied this method to determine the abundances of 21 elements. Our determinations are displayed in Figure 3 with error bars, and we compare our results to the abundances previously found by Adelman and al.(2003) who used the same effective temperature and surface gravity. 
Indeed, the parameters chosen by Adelman and al. differ as 
$\Delta T_{\textsl{eff}}=\pm250$K and $\Delta \log(g)=\pm0.25$ from our parameters. Furthermore, we used in our model atmosphere a microturbulence velocity up to $\xi_T=2.64~\pm0.66~km/s$ while Adelman and al. used $\xi_T=2.3~km/s$.

Our abundance analysis yields a pronounced underabundance of Sc and slight underabundances in C, O, Mg, Si and Ca, mild overabundances of the iron-peak elements and large overabundances of the rare-earth elements.
We find abundances which are consistent with the determinations of Adelman and al. (2003) except for Scandium. Our results differ from 0.01 dex up to 0.4 dex as we adopted new atomic data. 
All these new abundance determinations confirm the Am status for 68 Tauri.

\begin{center}
\resizebox{0.78\textwidth}{!}{
\setlength{\unitlength}{0.240900pt}
\ifx\plotpoint\undefined\newsavebox{\plotpoint}\fi
\sbox{\plotpoint}{\rule[-0.200pt]{0.400pt}{0.400pt}}%
\begin{picture}(1500,900)(0,0)
\sbox{\plotpoint}{\rule[-0.200pt]{0.400pt}{0.400pt}}%
\put(130.0,82.0){\rule[-0.200pt]{4.818pt}{0.400pt}}
\put(110,82){\makebox(0,0)[r]{$-2.5$}}
\put(1419.0,82.0){\rule[-0.200pt]{4.818pt}{0.400pt}}
\put(130.0,168.0){\rule[-0.200pt]{4.818pt}{0.400pt}}
\put(110,168){\makebox(0,0)[r]{$-2$}}
\put(1419.0,168.0){\rule[-0.200pt]{4.818pt}{0.400pt}}
\put(130.0,255.0){\rule[-0.200pt]{4.818pt}{0.400pt}}
\put(110,255){\makebox(0,0)[r]{$-1.5$}}
\put(1419.0,255.0){\rule[-0.200pt]{4.818pt}{0.400pt}}
\put(130.0,341.0){\rule[-0.200pt]{4.818pt}{0.400pt}}
\put(110,341){\makebox(0,0)[r]{$-1$}}
\put(1419.0,341.0){\rule[-0.200pt]{4.818pt}{0.400pt}}
\put(130.0,427.0){\rule[-0.200pt]{4.818pt}{0.400pt}}
\put(110,427){\makebox(0,0)[r]{$-0.5$}}
\put(1419.0,427.0){\rule[-0.200pt]{4.818pt}{0.400pt}}
\put(130.0,514.0){\rule[-0.200pt]{4.818pt}{0.400pt}}
\put(110,514){\makebox(0,0)[r]{$0$}}
\put(1419.0,514.0){\rule[-0.200pt]{4.818pt}{0.400pt}}
\put(130.0,600.0){\rule[-0.200pt]{4.818pt}{0.400pt}}
\put(110,600){\makebox(0,0)[r]{$0.5$}}
\put(1419.0,600.0){\rule[-0.200pt]{4.818pt}{0.400pt}}
\put(130.0,686.0){\rule[-0.200pt]{4.818pt}{0.400pt}}
\put(110,686){\makebox(0,0)[r]{$1$}}
\put(1419.0,686.0){\rule[-0.200pt]{4.818pt}{0.400pt}}
\put(130.0,773.0){\rule[-0.200pt]{4.818pt}{0.400pt}}
\put(110,773){\makebox(0,0)[r]{$1.5$}}
\put(1419.0,773.0){\rule[-0.200pt]{4.818pt}{0.400pt}}
\put(130.0,859.0){\rule[-0.200pt]{4.818pt}{0.400pt}}
\put(110,859){\makebox(0,0)[r]{$2$}}
\put(1419.0,859.0){\rule[-0.200pt]{4.818pt}{0.400pt}}
\put(143.0,82.0){\rule[-0.200pt]{0.400pt}{4.818pt}}
\put(143,41){\makebox(0,0){C}}
\put(143.0,839.0){\rule[-0.200pt]{0.400pt}{4.818pt}}
\put(207.0,82.0){\rule[-0.200pt]{0.400pt}{4.818pt}}
\put(207,41){\makebox(0,0){O}}
\put(207.0,839.0){\rule[-0.200pt]{0.400pt}{4.818pt}}
\put(271.0,82.0){\rule[-0.200pt]{0.400pt}{4.818pt}}
\put(271,41){\makebox(0,0){Mg}}
\put(271.0,839.0){\rule[-0.200pt]{0.400pt}{4.818pt}}
\put(335.0,82.0){\rule[-0.200pt]{0.400pt}{4.818pt}}
\put(335,41){\makebox(0,0){Si}}
\put(335.0,839.0){\rule[-0.200pt]{0.400pt}{4.818pt}}
\put(400.0,82.0){\rule[-0.200pt]{0.400pt}{4.818pt}}
\put(400,41){\makebox(0,0){Ca}}
\put(400.0,839.0){\rule[-0.200pt]{0.400pt}{4.818pt}}
\put(464.0,82.0){\rule[-0.200pt]{0.400pt}{4.818pt}}
\put(464,41){\makebox(0,0){Sc}}
\put(464.0,839.0){\rule[-0.200pt]{0.400pt}{4.818pt}}
\put(528.0,82.0){\rule[-0.200pt]{0.400pt}{4.818pt}}
\put(528,41){\makebox(0,0){Ti}}
\put(528.0,839.0){\rule[-0.200pt]{0.400pt}{4.818pt}}
\put(592.0,82.0){\rule[-0.200pt]{0.400pt}{4.818pt}}
\put(592,41){\makebox(0,0){V}}
\put(592.0,839.0){\rule[-0.200pt]{0.400pt}{4.818pt}}
\put(656.0,82.0){\rule[-0.200pt]{0.400pt}{4.818pt}}
\put(656,41){\makebox(0,0){Cr}}
\put(656.0,839.0){\rule[-0.200pt]{0.400pt}{4.818pt}}
\put(720.0,82.0){\rule[-0.200pt]{0.400pt}{4.818pt}}
\put(720,41){\makebox(0,0){Fe}}
\put(720.0,839.0){\rule[-0.200pt]{0.400pt}{4.818pt}}
\put(785.0,82.0){\rule[-0.200pt]{0.400pt}{4.818pt}}
\put(785,41){\makebox(0,0){Co}}
\put(785.0,839.0){\rule[-0.200pt]{0.400pt}{4.818pt}}
\put(849.0,82.0){\rule[-0.200pt]{0.400pt}{4.818pt}}
\put(849,41){\makebox(0,0){Ni}}
\put(849.0,839.0){\rule[-0.200pt]{0.400pt}{4.818pt}}
\put(913.0,82.0){\rule[-0.200pt]{0.400pt}{4.818pt}}
\put(913,41){\makebox(0,0){Sr}}
\put(913.0,839.0){\rule[-0.200pt]{0.400pt}{4.818pt}}
\put(977.0,82.0){\rule[-0.200pt]{0.400pt}{4.818pt}}
\put(977,41){\makebox(0,0){Y}}
\put(977.0,839.0){\rule[-0.200pt]{0.400pt}{4.818pt}}
\put(1041.0,82.0){\rule[-0.200pt]{0.400pt}{4.818pt}}
\put(1041,41){\makebox(0,0){Zr}}
\put(1041.0,839.0){\rule[-0.200pt]{0.400pt}{4.818pt}}
\put(1105.0,82.0){\rule[-0.200pt]{0.400pt}{4.818pt}}
\put(1105,41){\makebox(0,0){Ba}}
\put(1105.0,839.0){\rule[-0.200pt]{0.400pt}{4.818pt}}
\put(1170.0,82.0){\rule[-0.200pt]{0.400pt}{4.818pt}}
\put(1170,41){\makebox(0,0){La}}
\put(1170.0,839.0){\rule[-0.200pt]{0.400pt}{4.818pt}}
\put(1234.0,82.0){\rule[-0.200pt]{0.400pt}{4.818pt}}
\put(1234,41){\makebox(0,0){Ce}}
\put(1234.0,839.0){\rule[-0.200pt]{0.400pt}{4.818pt}}
\put(1298.0,82.0){\rule[-0.200pt]{0.400pt}{4.818pt}}
\put(1298,41){\makebox(0,0){Sm}}
\put(1298.0,839.0){\rule[-0.200pt]{0.400pt}{4.818pt}}
\put(1362.0,82.0){\rule[-0.200pt]{0.400pt}{4.818pt}}
\put(1362,41){\makebox(0,0){Dy}}
\put(1362.0,839.0){\rule[-0.200pt]{0.400pt}{4.818pt}}
\put(1426.0,82.0){\rule[-0.200pt]{0.400pt}{4.818pt}}
\put(1426,41){\makebox(0,0){Er}}
\put(1426.0,839.0){\rule[-0.200pt]{0.400pt}{4.818pt}}
\put(130.0,82.0){\rule[-0.200pt]{0.400pt}{187.179pt}}
\put(130.0,82.0){\rule[-0.200pt]{315.338pt}{0.400pt}}
\put(1439.0,82.0){\rule[-0.200pt]{0.400pt}{187.179pt}}
\put(130.0,859.0){\rule[-0.200pt]{315.338pt}{0.400pt}}
\put(0,510){\makebox(0,0)[r]{[X/H]}}
\put(143.0,463.0){\rule[-0.200pt]{0.400pt}{20.717pt}}
\put(133.0,463.0){\rule[-0.200pt]{4.818pt}{0.400pt}}
\put(133.0,549.0){\rule[-0.200pt]{4.818pt}{0.400pt}}
\put(207.0,433.0){\rule[-0.200pt]{0.400pt}{20.717pt}}
\put(197.0,433.0){\rule[-0.200pt]{4.818pt}{0.400pt}}
\put(197.0,519.0){\rule[-0.200pt]{4.818pt}{0.400pt}}
\put(271.0,458.0){\rule[-0.200pt]{0.400pt}{20.958pt}}
\put(261.0,458.0){\rule[-0.200pt]{4.818pt}{0.400pt}}
\put(261.0,545.0){\rule[-0.200pt]{4.818pt}{0.400pt}}
\put(335.0,463.0){\rule[-0.200pt]{0.400pt}{20.717pt}}
\put(325.0,463.0){\rule[-0.200pt]{4.818pt}{0.400pt}}
\put(325.0,549.0){\rule[-0.200pt]{4.818pt}{0.400pt}}
\put(400.0,445.0){\rule[-0.200pt]{0.400pt}{20.717pt}}
\put(390.0,445.0){\rule[-0.200pt]{4.818pt}{0.400pt}}
\put(390.0,531.0){\rule[-0.200pt]{4.818pt}{0.400pt}}
\put(464.0,113.0){\rule[-0.200pt]{0.400pt}{20.717pt}}
\put(454.0,113.0){\rule[-0.200pt]{4.818pt}{0.400pt}}
\put(454.0,199.0){\rule[-0.200pt]{4.818pt}{0.400pt}}
\put(528.0,536.0){\rule[-0.200pt]{0.400pt}{20.717pt}}
\put(518.0,536.0){\rule[-0.200pt]{4.818pt}{0.400pt}}
\put(518.0,622.0){\rule[-0.200pt]{4.818pt}{0.400pt}}
\put(592.0,579.0){\rule[-0.200pt]{0.400pt}{20.958pt}}
\put(582.0,579.0){\rule[-0.200pt]{4.818pt}{0.400pt}}
\put(582.0,666.0){\rule[-0.200pt]{4.818pt}{0.400pt}}
\put(656.0,567.0){\rule[-0.200pt]{0.400pt}{20.958pt}}
\put(646.0,567.0){\rule[-0.200pt]{4.818pt}{0.400pt}}
\put(646.0,654.0){\rule[-0.200pt]{4.818pt}{0.400pt}}
\put(720.0,524.0){\rule[-0.200pt]{0.400pt}{20.717pt}}
\put(710.0,524.0){\rule[-0.200pt]{4.818pt}{0.400pt}}
\put(710.0,610.0){\rule[-0.200pt]{4.818pt}{0.400pt}}
\put(785.0,643.0){\rule[-0.200pt]{0.400pt}{20.958pt}}
\put(775.0,643.0){\rule[-0.200pt]{4.818pt}{0.400pt}}
\put(775.0,730.0){\rule[-0.200pt]{4.818pt}{0.400pt}}
\put(849.0,616.0){\rule[-0.200pt]{0.400pt}{20.717pt}}
\put(839.0,616.0){\rule[-0.200pt]{4.818pt}{0.400pt}}
\put(839.0,702.0){\rule[-0.200pt]{4.818pt}{0.400pt}}
\put(913.0,619.0){\rule[-0.200pt]{0.400pt}{20.717pt}}
\put(903.0,619.0){\rule[-0.200pt]{4.818pt}{0.400pt}}
\put(903.0,705.0){\rule[-0.200pt]{4.818pt}{0.400pt}}
\put(977.0,616.0){\rule[-0.200pt]{0.400pt}{20.717pt}}
\put(967.0,616.0){\rule[-0.200pt]{4.818pt}{0.400pt}}
\put(967.0,702.0){\rule[-0.200pt]{4.818pt}{0.400pt}}
\put(1041.0,609.0){\rule[-0.200pt]{0.400pt}{20.717pt}}
\put(1031.0,609.0){\rule[-0.200pt]{4.818pt}{0.400pt}}
\put(1031.0,695.0){\rule[-0.200pt]{4.818pt}{0.400pt}}
\put(1105.0,683.0){\rule[-0.200pt]{0.400pt}{20.717pt}}
\put(1095.0,683.0){\rule[-0.200pt]{4.818pt}{0.400pt}}
\put(1095.0,769.0){\rule[-0.200pt]{4.818pt}{0.400pt}}
\put(1170.0,698.0){\rule[-0.200pt]{0.400pt}{20.958pt}}
\put(1160.0,698.0){\rule[-0.200pt]{4.818pt}{0.400pt}}
\put(1160.0,785.0){\rule[-0.200pt]{4.818pt}{0.400pt}}
\put(1234.0,674.0){\rule[-0.200pt]{0.400pt}{20.958pt}}
\put(1224.0,674.0){\rule[-0.200pt]{4.818pt}{0.400pt}}
\put(1224.0,761.0){\rule[-0.200pt]{4.818pt}{0.400pt}}
\put(1298.0,681.0){\rule[-0.200pt]{0.400pt}{20.717pt}}
\put(1288.0,681.0){\rule[-0.200pt]{4.818pt}{0.400pt}}
\put(1288.0,767.0){\rule[-0.200pt]{4.818pt}{0.400pt}}
\put(1362.0,654.0){\rule[-0.200pt]{0.400pt}{20.717pt}}
\put(1352.0,654.0){\rule[-0.200pt]{4.818pt}{0.400pt}}
\put(1352.0,740.0){\rule[-0.200pt]{4.818pt}{0.400pt}}
\put(1426.0,626.0){\rule[-0.200pt]{0.400pt}{20.717pt}}
\put(1416.0,626.0){\rule[-0.200pt]{4.818pt}{0.400pt}}
\put(1416.0,712.0){\rule[-0.200pt]{4.818pt}{0.400pt}}
\put(143,506){\makebox(0,0){$\bullet$}}
\put(207,476){\makebox(0,0){$\bullet$}}
\put(271,501){\makebox(0,0){$\bullet$}}
\put(335,506){\makebox(0,0){$\bullet$}}
\put(400,488){\makebox(0,0){$\bullet$}}
\put(464,156){\makebox(0,0){$\bullet$}}
\put(528,579){\makebox(0,0){$\bullet$}}
\put(592,622){\makebox(0,0){$\bullet$}}
\put(656,610){\makebox(0,0){$\bullet$}}
\put(720,567){\makebox(0,0){$\bullet$}}
\put(785,686){\makebox(0,0){$\bullet$}}
\put(849,659){\makebox(0,0){$\bullet$}}
\put(913,662){\makebox(0,0){$\bullet$}}
\put(977,659){\makebox(0,0){$\bullet$}}
\put(1041,652){\makebox(0,0){$\bullet$}}
\put(1105,726){\makebox(0,0){$\bullet$}}
\put(1170,742){\makebox(0,0){$\bullet$}}
\put(1234,717){\makebox(0,0){$\bullet$}}
\put(1298,724){\makebox(0,0){$\bullet$}}
\put(1362,697){\makebox(0,0){$\bullet$}}
\put(1426,669){\makebox(0,0){$\bullet$}}
\put(143,414){\makebox(0,0){$\circ$}}
\put(207,462){\makebox(0,0){$\circ$}}
\put(271,498){\makebox(0,0){$\circ$}}
\put(335,505){\makebox(0,0){$\circ$}}
\put(400,462){\makebox(0,0){$\circ$}}
\put(464,338){\makebox(0,0){$\circ$}}
\put(528,555){\makebox(0,0){$\circ$}}
\put(592,631){\makebox(0,0){$\circ$}}
\put(656,598){\makebox(0,0){$\circ$}}
\put(720,572){\makebox(0,0){$\circ$}}
\put(785,610){\makebox(0,0){$\circ$}}
\put(849,652){\makebox(0,0){$\circ$}}
\put(913,704){\makebox(0,0){$\circ$}}
\put(977,638){\makebox(0,0){$\circ$}}
\put(1041,679){\makebox(0,0){$\circ$}}
\put(1105,724){\makebox(0,0){$\circ$}}
\put(1170,643){\makebox(0,0){$\circ$}}
\put(1234,695){\makebox(0,0){$\circ$}}
\put(1298,799){\makebox(0,0){$\circ$}}
\put(1362,702){\makebox(0,0){$\circ$}}
\put(143,514){\usebox{\plotpoint}}
\put(143.00,514.00){\usebox{\plotpoint}}
\put(163.76,514.00){\usebox{\plotpoint}}
\put(184.51,514.00){\usebox{\plotpoint}}
\put(205.27,514.00){\usebox{\plotpoint}}
\put(226.02,514.00){\usebox{\plotpoint}}
\put(246.78,514.00){\usebox{\plotpoint}}
\put(267.53,514.00){\usebox{\plotpoint}}
\put(288.29,514.00){\usebox{\plotpoint}}
\put(309.04,514.00){\usebox{\plotpoint}}
\put(329.80,514.00){\usebox{\plotpoint}}
\put(350.55,514.00){\usebox{\plotpoint}}
\put(371.31,514.00){\usebox{\plotpoint}}
\put(392.07,514.00){\usebox{\plotpoint}}
\put(412.82,514.00){\usebox{\plotpoint}}
\put(433.58,514.00){\usebox{\plotpoint}}
\put(454.33,514.00){\usebox{\plotpoint}}
\put(475.09,514.00){\usebox{\plotpoint}}
\put(495.84,514.00){\usebox{\plotpoint}}
\put(516.60,514.00){\usebox{\plotpoint}}
\put(537.35,514.00){\usebox{\plotpoint}}
\put(558.11,514.00){\usebox{\plotpoint}}
\put(578.87,514.00){\usebox{\plotpoint}}
\put(599.62,514.00){\usebox{\plotpoint}}
\put(620.38,514.00){\usebox{\plotpoint}}
\put(641.13,514.00){\usebox{\plotpoint}}
\put(661.89,514.00){\usebox{\plotpoint}}
\put(682.64,514.00){\usebox{\plotpoint}}
\put(703.40,514.00){\usebox{\plotpoint}}
\put(724.15,514.00){\usebox{\plotpoint}}
\put(744.91,514.00){\usebox{\plotpoint}}
\put(765.66,514.00){\usebox{\plotpoint}}
\put(786.42,514.00){\usebox{\plotpoint}}
\put(807.18,514.00){\usebox{\plotpoint}}
\put(827.93,514.00){\usebox{\plotpoint}}
\put(848.69,514.00){\usebox{\plotpoint}}
\put(869.44,514.00){\usebox{\plotpoint}}
\put(890.20,514.00){\usebox{\plotpoint}}
\put(910.95,514.00){\usebox{\plotpoint}}
\put(931.71,514.00){\usebox{\plotpoint}}
\put(952.46,514.00){\usebox{\plotpoint}}
\put(973.22,514.00){\usebox{\plotpoint}}
\put(993.98,514.00){\usebox{\plotpoint}}
\put(1014.73,514.00){\usebox{\plotpoint}}
\put(1035.49,514.00){\usebox{\plotpoint}}
\put(1056.24,514.00){\usebox{\plotpoint}}
\put(1077.00,514.00){\usebox{\plotpoint}}
\put(1097.75,514.00){\usebox{\plotpoint}}
\put(1118.51,514.00){\usebox{\plotpoint}}
\put(1139.26,514.00){\usebox{\plotpoint}}
\put(1160.02,514.00){\usebox{\plotpoint}}
\put(1180.77,514.00){\usebox{\plotpoint}}
\put(1201.53,514.00){\usebox{\plotpoint}}
\put(1222.29,514.00){\usebox{\plotpoint}}
\put(1243.04,514.00){\usebox{\plotpoint}}
\put(1263.80,514.00){\usebox{\plotpoint}}
\put(1284.55,514.00){\usebox{\plotpoint}}
\put(1305.31,514.00){\usebox{\plotpoint}}
\put(1326.06,514.00){\usebox{\plotpoint}}
\put(1346.82,514.00){\usebox{\plotpoint}}
\put(1367.57,514.00){\usebox{\plotpoint}}
\put(1388.33,514.00){\usebox{\plotpoint}}
\put(1409.09,514.00){\usebox{\plotpoint}}
\put(1426,514){\usebox{\plotpoint}}
\sbox{\plotpoint}{\rule[-0.200pt]{0.400pt}{0.400pt}}%
\put(130.0,82.0){\rule[-0.200pt]{0.400pt}{187.179pt}}
\put(130.0,82.0){\rule[-0.200pt]{315.338pt}{0.400pt}}
\put(1439.0,82.0){\rule[-0.200pt]{0.400pt}{187.179pt}}
\put(130.0,859.0){\rule[-0.200pt]{315.338pt}{0.400pt}}
\end{picture}
}
\captionof{figure}{The found abundance pattern in 68 Tauri:\textit{ circles (Adelman and al.), dots (this work).}\\As usual, the script [X/H] means $log(X/H)_*-log(X/H)_{\odot}$ with the solar abundances recommanded by Grevesse and Sauval(1998) in the usual scale where $log N(H)=+12.00$.}
\end{center}

\section{Conclusions}
 
Our abundance analysis yields a pronounced underabundance of Sc and slight underabundances in C, O, Mg, Si and Ca, mild overabundances of the iron-peak elements and large overabundances of the rare-earth elements. All these new abundance determinations confirm the Am status for 68 Tauri.
Thanks to the improvement of atomic data, we have enlarged and improved the elemental abundances of 68 Tauri. The new results on the rare-earth group confirm the Am 
peculiarity of 68 Tauri. The inclusion of the hyperfine structure of the various isotopes of Ba II leads us to decrease the baryum abundance in 68 Tauri.
We stress the importance of taking into account the Hyperfine Structure for all isotopes when available in order to derive accurate abundances.

\nocite{1985MNRAS.217..305M}\nocite{1992IAUS..149..225K}\nocite{1998AJ....115.1640M}\nocite{2003A&A...406..987P}\nocite{1998SSRv...85..161G}
\bibliographystyle{aa}  
\bibliography{martinet} 

\end{document}